\documentstyle[12pt]{article} 
\newcommand{\cftnote} {\renewcommand{\thefootnote}{\fnsymbol{footnote}}}
\newcommand{\resetftnote}{\setcounter{footnote}{0}}

\newcommand{\be}{\begin{equation}} \newcommand{\ee}{\end{equation}}
\newcommand{\bea}{\begin{eqnarray}} \newcommand{\eea}{\end{eqnarray}}

\def\bm#1{\mbox{\boldmath{$#1$}}}

\def\IR{\relax{\rm I\kern-.18em R}}
\def\IN{\relax{\rm I\kern-.18em N}}
\def\l{\lambda}
\def\z{\zeta}
\def\r{\rho}
\font\cmss=cmss10 \font\cmsss=cmss10 at 7pt
\def\IZ{\relax\ifmmode\mathchoice
{\hbox{\cmss Z\kern-.4em Z}}{\hbox{\cmss Z\kern-.4em Z}}
{\lower.9pt\hbox{\cmsss Z\kern-.4em Z}}
{\lower1.2pt\hbox{\cmsss Z\kern-.4em Z}}\else{\cmss Z\kern-.4em Z}\fi}

\def\IR{\relax{\rm I\kern-.18em R}}

\begin{document}

\begin{flushright} FFUOV-98-04\\ hep-th/9804025\\ March 1998 
\end{flushright}

\cftnote

\vspace{1mm} {\Large \bf \begin{center} 
On thermodynamic approaches to\\ conformal field theory
 \end{center} } \vspace{2mm} 
\setcounter{footnote}{3}
\begin{center} 
{\large Jos{\'e} Gaite%
\footnote{E-mail address: gaite@laeff.esa.es and gaite@fisi77.ciencias.uniovi.es}} \\[4mm] 
{\it Laboratorio de Astrof\'{\i}sica Espacial y F\'{\i}sica Fundamental,\\ 
Apartado 50727, 28080 Madrid, Spain\\and\\
Departamento de F\'{ \i}sica,
Facultad de Ciencias,\\
Universidad de Oviedo,
c/ Calvo Sotelo s.n.,
33007 OVIEDO.}\\ 
\end{center} 
\resetftnote

\vspace{2mm}

\begin{abstract}
We present the thermodynamic Bethe ansatz as a way to 
factorize the partition function of 
a 2d field theory, in particular, a conformal field theory and 
we compare it with another 
approach to factorization due to K. Schoutens which consists of diagonalizing 
matrix recursion relations 
between the partition functions at consecutive levels. We prove that 
both are equivalent, taking as examples the $SU(2)$ spinons and 
the 3-state Potts model. In the latter case we see that there are 
two different thermodynamic Bethe ansatz equation systems with 
the same physical content, of which 
the second is new and corresponds to a one-quasiparticle representation, 
as opposed to the usual two-quasiparticle representation. This new 
thermodynamic Bethe ansatz system leads to a new dilogarithmic formula for 
the central charge of that model.
\end{abstract}

\global\parskip 4pt

\vspace{1cm}


\vspace{1cm}

\newpage

\section{Introduction}

The representation of conformal field theories (CFT's) 
in terms of quasi-particles is gaining 
importance, especially in regard to their application in statistical or 
condensed matter physics \cite{DKKMM,BerPaSer}. In this representation the 
Hilbert space is built up from free particle states, in a similar way to 
the description of the Ising model in terms of free fermions. 
It is remarkable that the rules for constructing the eigenvalue spectra can 
generally be interpreted as a realization of Haldane exclusion statistics
\cite{Haldane,Wu}. Thus this representation is usually called fermionic or 
quasi-fermionic. The partition function of a particular CFT is easily derived 
as a combinatorial sum, once the energy is expressed as a sum of 
single particle levels of quasi-fermionic type.

On the other hand, the powerful methods of CFT based on free field representations 
provide expresions for the conformal characters and 
hence for the partition function in the form of bosonic sums \cite{DKKMM}. 
Here one may question whether the quasi-particle
representation contributes something, apart from the connections it implies. 
The answer lies in the fact, sometimes not sufficiently appreciated, 
that even though the partition function contains all the information on the CFT, 
this information is usually very hard to extract. 
The reason is that the expression of the partition function as a combinatorial sum 
is of little utility to obtain thermodynamic quantities. For this one needs 
the free energy, that is, the logarithm of the partition function, and its 
derivatives. In a free theory (take for example the fermionic representation of 
the Ising model) the summations in the (grand) partition function can be made, 
obtaining the usual infinite product expression. Its logarithm has 
a nice expression as an infinite sum, from which one derives 
the relevant thermodynamic quantities. Consequently, a sensible way to derive 
the thermodynamics of a CFT is to express the partition function in 
a factorized form. However, this is seldom achievable exactly.%
\footnote{For free CFT, bosonic or fermionic with periodic or twisted boundary 
conditions, it is a standard exercise in CFT. It admits a mathematical 
formulation as the factorization of Jacobi elliptic theta functions as a 
consequence of the Gauss-Jacobi triple product identity.}
In some cases one can achieve an approximate factorization nevertheless.

It is in this context where the quasi-particle representation shows its 
utility. These quasi-particles are free except for the constraints embodied 
by their statistics or, in other words, the rules for the filling of 
single-particle levels. They turn out to be non trivial, unlike those for bosons 
or pure fermions, and constitute what has been called statistical interaction. 
Nevertheless, in the thermodynamic limit the entropy can be expressed as a sum, 
and the partition function as well, in a way that generalizes the derivation of 
the Bose or Fermi distributions in the microcanonical ensemble \cite{Wu,Isakov}. 
It turns out that the equations for the occupation numbers $n_i$ are deduced 
from conditions which adopt the form 
of thermodynamic Bethe ansatz (TBA) equations for massless particles. 
This should come as no surprise 
for the TBA is essentially the same kind of approach: Starting with a system 
which can be described by the elastic scattering of quasi-particles 
(Bethe ansatz) the thermodynamic approach \cite{YY} consists of finding an 
expression for their entropy and hence deriving the distribution 
of quasi-particles.

A new method for factorizing the partition function of a CFT has been proposed 
recently by K.\ Schoutens \cite{Schou}. He calls it ``transfer matrix for 
truncated chiral spectra" and consists of finding recursion relations between 
partition functions truncated at some level in the spectrum. Typically, 
he finds that only a small number of truncated partition functions enter in 
the recursion relations, which can be encoded in a matrix form. Therefore, 
the partition function at a generic level is given by the product of the 
succesive recursion matrices. One can then diagonalize the matrix and 
substitute the partition function by the product of the succesive 
largest eigenvalues, putting it in a factorized form. However, the eigenvalues 
take rather complicated expressions even in the simpler cases. A useful 
simplification is to consider the relevant levels sufficiently high and 
hence to take the limit $l \rightarrow \infty$ while keeping $l/L$ finite, 
with $l$ and 
$L$ being the level number and the system size, respectively. This is again 
the thermodynamic limit. Schoutens' method is thus in spirit closely related 
with the TBA. However, he asserts that the distributions he obtains are 
different from those obtained from TBA equations and 
says that the relation between the two 
approaches is not clear. We shall show here that, in addition to being 
conceptually analogous, these two approaches lead to the same equations. 
We shall also show that their comparison reveals new TBA systems of 
equations for known models.
 
The factorization properties of CFT and their connection with the TBA approach 
have been considered in \cite{ByFri} from a different point of view. They 
consider characters which factorize {\em exactly} and show that one can 
take advantage of it to calculate the central charge. They further show that 
a saddle point approximation of the characters, which is correct in 
the thermodynamic limit, leads to the TBA equations (see also \cite{NaReTer}). 
However, they make no direct connection between the TBA equations and factorization 
nor do they attempt to connect with Schoutens' approach.

\section{A simple case: $SU(2)_1$ and semions}

The simplest recursion relation occurs for the trivial case, 
massless free fermions. The one-fermion energy levels are 
$\epsilon_l = (l+{\frac{1}{2}})\,{\pi \over L}$. 
Let us denote by $Z_l(q,z)$, with $q=e^{-\beta\frac{\pi}{L}}$ and 
the fugacity $z=e^{\beta\,\mu}$,
the partition function at level $l$.\footnote{We consider the chiral or 
holomorphic part of the partition function, being understood that 
the total partition function is a product of the holomorphic part and 
the antiholomorphic part.} The addition of the level $l+1$ provides 
a new state to place a fermion so
\be
Z_{l+1}(q,z) = Z_l(q,z) + q^{l+{\frac{3}{2}}}\,z\,Z_l(q,z) = 
(1+ q^{l+{\frac{1}{2}}}\,z)\,Z_l(q,z).
\ee
This recursion relation immediately gives the factorized form 
$Z_l(q,z) = \prod_{j=0}^{l}(1+ q^{j+{\frac{1}{2}}}\,z)$, which is exact. 
As is well known, the central charge can be computed as 
the leading finite-size correction to the free energy
$$
-\beta\,F(\beta,\mu) = \sum_{j=0}^{l} \log\left[{1 + q^{j+{\frac{1}{2}}}\,z}\right]
$$
for $\mu=0$, namely, 
\be
{\frac{\pi}{6\,\beta}}\,c = -\beta\,{F(\beta,0)\over L} = 
\int_{-\infty}^\infty {dp\over 2\pi} 
\ln\left(1+ e^{-\beta\,|p|}\right).
\ee
It yields $c=\frac{1}{2}$. 

Apart from   
this trivial case, the simplest case is that of semions, namely, particles 
with statistical parameter $g=1/2$ and double maximum occupation per level. 
They appear as the elementary excitations of a spin chain with long range 
interactions \cite{Hal,Shas}, whose conformal limit is described by 
the $SU(2)_1$ WZW model. In regard to that connection they are called 
spinons and exhibit a particular symmetry, the Yangian symmetry, of 
the WZW model \cite{BerPaSer,BoLuSa}. There are two types of spinons, 
according to the polarization of their spin, usually denoted with 
a $+$ or $-$ script. If we consider only one type, $+$ say, we have 
a system of pure semions. Their TBA equations are a particular case of 
those of the ideal gas of particles with fractional statistics,
\bea
\label{su2TBA}
w(\zeta)^g\,[1+w(\zeta)]^{1-g} = \zeta^{-1} \equiv e^{\beta\,(\epsilon - \mu)},\\
n_i = {1 \over w(\zeta_i)+g}
\eea
for $g=1/2$ \cite{Wu}. They have the simple solution 
\be
n_i = {1 \over \sqrt{1/4 + \exp[2\,\beta\,(\epsilon_i - \mu)]}}.
\ee
Furthermore, the free energy is given by the sum over one-particle levels
\be
F = -T \sum_{i} \log\left[1+w(\zeta_i)^{-1}\right],   \label{Fsu2}
\ee
with the corresponding $w(\z_i)$.

The spinon partition function is  
\be
Z(q;z) = \sum_{n^+,\,n^-} z_+^{n^+} \,z_-^{n^-}\,  
\frac{q^{(n^++n^-)^2/4}}{(q)_{n^+}\,(q)_{n^-}}         \label{ZSU2}
\ee
with $(q)_n= \prod_{j=1}^n(1-q^j)$ \cite{BerPaSer}. There are two independent 
characters included in it, namely, the vacuum character and the character 
corresponding to the $SU(2)$ fundamental representation. They are obtained by 
restricting the sum to $n^++n^-$ even or odd, respectively. 
One immediately deduces from (\ref{ZSU2}) that only integer powers of $q$ appear 
in the first one whereas the powers of $q$ in 
the second one are integer modulo 1/4. 
It is convenient to conventionally split every level $l$ into an odd sublevel 
$l+1/2$ and an even sublevel $l+3/2$ \cite{Schou}. Then one concludes that 
the vacuum character correspond to a highest occupied odd level whereas the 
other one correspond to a highest occupied even level. Let us denote them by 
$Z_0 = \chi^0(q,z_+,z_-)$ and $Z_1 = \chi^1(q,z_+,z_-)$, respectively.

The partition function for + spinons is obtained by setting 
the fugacity $z_- = 0$ in the total partition function (\ref{ZSU2}).
By examining the filling 
of levels it is possible to find a recursion relation for the truncated even and 
odd partition functions:
\be
\left( \begin{array}{c} (Z_1)_l \\ (Z_0)_l \end{array} \right)
=
\left( \begin{array}{cc} 
                  1  & 
                  q^{l-{\frac{3}{4}}} \, z_+  \\
                  q^{l-{\frac{1}{4}}} \, z_+  & 
                  1 + q^{2\,l-1} \, z_+^2 
       \end{array} \right)
\left( \begin{array}{c} (Z_1)_{l-1} \\ (Z_0)_{l-1} \end{array} \right),
\label{recM1}
\ee
with initial values $Z_0 = 1$ and $Z_1 = 0$ \cite{Schou}. 
The partition function at level $l$ is given by the product of the 
succesive recursion matrices for $j=1,\dots,l$.
Taking after the transfer matrix method, Schoutens then proposes 
to approximate this exact expression by the product of 
the largest eigenvalues of the matrices. One can easily calculate the eigenvalues 
of the recursion matrix
\be
\left| \begin{array}{cc} 
                  1 - \lambda_j  & 
                  q^{j-{\frac{3}{4}}} \, z_+  \\
                  q^{j-{\frac{1}{4}}} \, z_+  & 
                  1 + q^{2\,j-1} \, z_+^2 - \lambda_j
       \end{array} \right| = 
1 + \left( -2 - {q^{-1 + 2\,j}}\,{z_+^2} \right) \lambda_j  + 
  {{\lambda_j}^2}=0.
\ee
Under the substitution 
${\lambda }\rightarrow {1 + w^{-1}}$
\be
1 - {q^{2\,j-1}}\,{z_+^2} \,w \left( 1 + w \right) = 0.
\ee
It coincides with (\ref{su2TBA}) for $g=1/2$ if we take $\epsilon_j = 
(j+{\frac{1}{2}})\,{\pi \over L}$. Notice that these energies exactly coincide 
with one-fermion levels.

The factorized partition function given by the product of the largest eigenvalues 
$$\lambda_l = 1 + {\frac{{q^{-1 + 2\,l}}\,{z_+^2}}{2}} + 
  {\frac{{q^{-1 + l}}\,z_+\,{\sqrt{4\,q + {q^{2\,l}}\,{z_+^2}}}}{2}}$$ 
also yields a free energy identical with (\ref{Fsu2}):
\be
\beta\,F = -\log\left[\prod_{j} \lambda_j \right] = 
\sum_{j} \log\left[{1 + w_j^{-1}}\right].
\ee
The central charge is 
\be
\,c = 
{\frac{6}{\pi^2}}
\int _{0}^{1}\frac{d\z}{\z}\,
{\log \left[{\frac{2 + {\z^2} + \z\,{\sqrt{4 + {\z^2}}}}{2}}\right]}
= {\frac{6}{\pi^2}}\,L\left(\frac{{\sqrt{5}}-1}{2}\right)
= \frac{3}{5},
\ee
with $L(x)$ being the Rogers dilogarithm function \cite{Lewin}.

The unpolarized case leads to a more complicated recursion relation 
\be
\left( \!\!\begin{array}{c} (Z_1)_l \\ (Z_0)_l \end{array} \!\!\right)
=
\left(\!\! \begin{array}{cc} 
           1-q^{2\,l-2}\,z_+\, z_- &  
           q^{l-{\frac{3}{4}}}  (z_+ + z_-) \\
           q^{l-{\frac{1}{4}}}  (z_+ + z_-) (1-q^{2\,l-2}\,z_+\, z_-) & 
           1 + q^{2\,l-1} (z_+^2 + z_+\, z_- + z_-^2)
       \end{array} \!\!\right)
\left(\!\! \begin{array}{c} (Z_1)_{l-1} \\ (Z_0)_{l-1} \end{array}\!\! \right)
\label{recM2}
\ee
and hence to a more complicated equation
\bea
1 - {q^{-2 + 2\,l}}\,z_+\, z_- - 
    {q^{-1 + 2\,l}}\,z_+\, z_- + 
    {q^{-3 + 4\,l}}\,{z_+^2}\,{z_-^2} - \nonumber\\ 
    \left( 2 + {q^{-1 + 2\,l}}\,{z_+^2} - 
       {q^{-2 + 2\,l}}\,z_+\, z_- + 
       {q^{-1 + 2\,l}}\,z_+\, z_- + 
       {q^{-1 + 2\,l}}\,{z_-^2} \right)
     \lambda  + {{\lambda }^2} = 0.
\label{eqM2}
\eea
We would like to identify this equation with some sort of TBA equation. Before 
delving into this question, let us notice that the solutions of 
Eq.\ (\ref{eqM2}) are quite unwieldy. However, 
when the chemical potentials of both polarizations are equal, the system 
behaves as two independent fermions with total central charge $c=1$, as 
corresponds to the $SU(2)_1$ WZW model \cite{Schou}. 
Nevertheless, there is no simplification in 
the solutions of (\ref{eqM2}) unless we take the thermodynamic limit 
$l \gg 1$. Then (\ref{eqM2}) adopts the form 
\be
{\lambda^2} 
- \lambda \left( 2 + {\zeta_+}^2 + {\zeta_-}^2 \right) +
{{\left(1 - \zeta_+ \,\zeta_- \right) }^2} = 0,
      \label{eigenSU2}
\ee
with $\zeta_\pm = {q^{l}\,z_\pm}$. In this limit, the solutions 
drastically simplify for $z_+ = z_-$, yielding
\be
\lambda = (1 \pm \zeta)^2.
\label{solSU2}
\ee
The larger solution (with the plus sign) leads to 
\be
\beta\,F = -\log\left[\prod_{j} \lambda_j \right] = 
-2\,\sum_{j} \log\left(1 + q^{j}\,z\right),
\ee
corresponding to two free fermions.

The TBA equations for unpolarized spinons can be written once 
the statistical matrix $g_{ij}$ is known \cite{Wu}. This matrix can be determined 
from the expression of the exponent of $q$ in the partition function 
(\ref{ZSU2}) to be 
$$
G = \left( \begin{array}{cc} {\frac{1}{2}} & {\frac{1}{2}} \\ 
{\frac{1}{2}} & {\frac{1}{2}} \end{array} \right).
$$
We obtain the TBA equations
\bea
{\frac{\left({\lambda_+} - 1 \right) \,
       {\sqrt{{\lambda_-}}}}{{\sqrt{{\lambda_+}}}}} = {\zeta_+},\\
{\frac{\left({\lambda_-} - 1\right) \,
       {\sqrt{{\lambda_+}}}}{{\sqrt{{\lambda_-}}}}} = {\zeta_-},
\label{TBASU2}
\eea
with the {\em definition} $\lambda_a = {1 + w_a^{-1}}$ for $a=+,-$. 
Let us remark that $\lambda_a$ are not eigenvalues and are not 
related to any recursion matrix.
The free energy is given by the sum over polarizations and over one-particle 
levels for each polarization
\be
\beta\,F = -\sum_{a,j} \log\left[1+w_a(\zeta_j)^{-1}\right] =
-\sum_{j} \log\left[{\lambda_+}\,{\lambda_-}\right]_j.   \label{FSU2}
\ee
Therefore, we are actually interested in the product ${\lambda_+}\,{\lambda_-}$ 
of the solutions of the TBA equations (\ref{TBASU2}) for given ${\zeta_+}$ and 
${\zeta_-}$ rather than in their independent value. We can add to these 
two equations a third one, $\lambda={\lambda_+}\,{\lambda_-}$ and 
eliminate the intermediate variables ${\lambda_+}$ and ${\lambda_-}$ to 
get an equation for $\lambda$. It turns out to be precisely the 
eigenvalue equation (\ref{eigenSU2}). Furthermore, the TBA free energy 
(\ref{FSU2}) corresponds to the factorized partition function being the product 
of eigenvalues $\prod_{j} \lambda_j ({\zeta_+},{\zeta_-})$. 

For $\mu_+ \neq \mu_-$ there is an interesting physical quantity which can be 
read out from the free energy, namely, the total magnetization, that is, 
the difference between the total number of up and down spinons $N_+ - N_-$ 
\cite{Schou}: 
\bea
N_+ - N_- = -{\partial F\over \partial \mu_+}+ {\partial F\over \partial \mu_-} 
= \int _{0}^{1}\frac{d\z}{\z}\,
{\frac{2\,\zeta \left( {z} - {z^{-1}} \right) }
   {{\sqrt{4 + {{\zeta }^2}\left( {z} - {z^{-1}} \right)^2}}}}
\nonumber\\ = 2\,{\rm ArcSinh}({\frac{{z} - {z^{-1}}}{2}}) = 2\,\log z = 
\beta\,(\mu_+ - \mu_-),    \label{magnet}
\eea
where we have introduced the variables 
$$z = \sqrt{\zeta_+/\zeta_-} = e^{\beta\,(\mu_+ - \mu_-)/2}$$ 
and $$\zeta = \sqrt{\zeta_+\,\zeta_-} = e^{-\beta\,[\epsilon-
(\mu_+ + \mu_-)/2]}.$$
Eq.\ (\ref{magnet}) expresses that the magnetization is proportional to 
the external magnetic field.

\section{The three-state Potts model and $\IZ_n$ parafermions}

The trivial case of massless real free fermions corresponds to the CFT of 
the critical Ising model. The natural (and oldest) generalization of this 
model consists of taking a site variable which can take three values 
instead of two, constituting the three-state Potts model. This model has 
been long known to be integrable and it has been long (but not as long) known 
to be describable in terms of particles with fractional statistics, which 
are generalizations of the Ising fermions and are called parafermions. 
A quasi-fermionic expression for the characters of the corresponding CFT 
is given by the Lepowski-Primc branching functions \cite{DKKMM,KKMM9211102}. 
This quasi-fermionic representation in terms of two conjugate parafermions 
carrying $\IZ_3$ charges +1 or $-1$ is not the only one possible 
\cite{DKMM93} but it is the correct one to describe the three-state Potts model 
at a non critical temperature as the perturbation produced by giving mass to 
the parafermions \cite{Zamo1,Zamo2}.  

The TBA equations for the three-state Potts model were obtained in \cite{Zamo2}. 
They can also be obtained from the Lepowski-Primc branching function 
representation of the characters, 
\be
q^{1/30}\,b_{2\,Q-l}^l(q) = q^{\frac{l(2-l)}{16}}\!\!\!
\sum_{m_1,m_2=0\atop m_1-m_2\equiv Q\ ({\rm mod~3})}^\infty \!\!
\frac{q^{{\bm m}\,C^{-1}_{A_2}\,{\bm m}^t+L_l({\bm m})}}{(q)_{m_1}(q)_{m_2}}~~,
\label{LepPrim}
\ee
where $$C_{A_2} = \left(\begin{array}{rr}2&-1\\-1&2\end{array}\right)$$
is the Cartan matrix of the Lie algebra $A_2$, and
$L_0({\bm m})=0$, $L_1({\bm m})=(2\,m_1+m_2)/3$,
$L_2({\bm m})=(m_1+2\,m_2)/3$.
It leads to the statistical matrix 
$$
G = 2\,C^{-1}_{A_2}=\left( \begin{array}{cc} {\frac{4}{3}} & {\frac{2}{3}} \\ 
{\frac{2}{3}} & {\frac{4}{3}} \end{array} \right).
$$
Hence, the TBA equations are: 
\bea
\label{TBA1Z3}
({\l_1} - 1)\,{\l_1}^{1/3}\,{\l_{\bar{1}}}^{2/3} = \z_1,   \\
\noalign{\vspace{0.5ex}}
\hspace{3.em} ({\l_{\bar{1}}} - 1)\,{\l_1}^{2/3}\,{\l_{\bar{1}}}^{1/3} = 
\z_{\bar{1}}.
\label{TBA2Z3}
\eea
Unlike the TBA equations in the previous section, these equations 
are very complicated. 
Their algebraic solution can only be obtained with the help of 
algebraic computing programs, such as Mathematica, and it is huge:
It occupies about 600 Kbytes of memory, that is, about 300 written pages! 
Moreover, it took 117847 seconds of CPU time to obtain it with a modern 
Pentium computer. Therefore, it is sensible to examine first particular cases.

The simplest particular case occurs for equal fugacities of positive and 
negative particles, $z_1 = z_{\bar{1}}$. Then it is like if 
the two equations are the same and we actually have only one,%
\footnote{Something similar happens for 
spinons described by the TBA equations (\ref{TBASU2}). In this case the 
single equation for $\zeta_+ = \zeta_-$ is trivial and immediately leads to 
the solution (\ref{solSU2}).}
\be
({\l} - 1)\,{\l} = \z.
\ee
The solution of this equation is elementary
\be
\l = \frac{1 \pm {\sqrt{1 + 4\,\z}}}{2}.
\label{solz3}
\ee
The central charge is 
\be
\,c = 
2\,{\frac{6}{\pi^2}}
\int _{0}^{1}\frac{d\z}{\z}\,
\log ({\frac{1 + {\sqrt{1 + 4\,\z}}}{2}}) 
= 2\,{\frac{6}{\pi^2}}\,L\left(\frac{3 - {\sqrt{5}}}{2}\right)
= \frac{4}{5}.
\label{cZ3}
\ee
Before proceeding to the other important particular case, let us remark that 
the algebraic expression obtained for the general solution is 
singular for $\zeta_1 = \zeta_{\bar{1}}$. The limit 
$\zeta_1 \rightarrow \zeta_{\bar{1}}$ coincides with (\ref{solz3}) if 
performed correctly, for the function is continuous there. 

The other obvious singularity of the expression for the general solution occurs 
for $\zeta_{\bar{1}} = \zeta_1^2$. The limit can again be performed and 
coincides with the solution of the equations (\ref{TBA1Z3}) and (\ref{TBA2Z3}) 
for $\zeta_{\bar{1}} = \zeta_1^2$. This solution is still quite hefty but 
far from the general solution: It takes about two pages. We may wonder to what 
physical situation this solution applies. According to the definition of $\z$, 
it corresponds to having $\mu_{\bar{1}} = 2\,\mu_1$ and 
$\epsilon_{\bar{1}} = 2\,\epsilon_1$. The first condition has a simple 
interpretation: The levels are filled asymmetrically for particles 1 and 
${\bar{1}}$, with an average number of the latter double than of the former. 
The second condition seems to imply an asymmetry in the one-particle levels, 
with those for particle ${\bar{1}}$ having double energy than those 
of particle 1. This would break the symmetry between both types of particles. 
However, there is an interpretation that encompasses both conditions and does 
not break this symmetry: The one-particle levels can still be the same if 
every particle ${\bar{1}}$ is considered as a composite of two particles 1. 
This interpretation actually leads to a different representation of the model 
in terms of only one type of particle which is interesting on its own. 
Moreover, it is in this representation 
that the recursion matrix has been obtained \cite{Schou}. It reads  
\be
R = \left( \begin{array}{ccc} 
           (1-\z^3)     & \z^2      &  \z     \\
           \z\,(1-\z^3)    &  1       & 2\,\z^2   \\
           2\,\z^2\,(1-\z^3) & \z\,(1+\z^3) & 1+2\,\z^3
       \end{array} \right) .
\label{recZ3}
\ee
Its eigenvalue equation is
\be
- {{\left( 1 - {{\zeta }^3} \right) }^3} + 
  \left( 3 - 3\,{{\zeta }^3} - {{\zeta }^6} \right) 
   \lambda  - \left( 3 + {{\zeta }^3} \right) 
   {{\lambda }^2} + {{\lambda }^3} = 0.
\label{eigenZ3}
\ee 
The solution of this equation is straightforward with the use of 
the Cardano formula for cubic equations. 

At this point, we would like to relate the restricted TBA equations for 
the representation by one type of particles,
\bea
\label{TBA1Z3r}
({\l_1} - 1)\,{\l_1}^{1/3}\,{\l_{\bar{1}}}^{2/3} = \z,  \\
\noalign{\vspace{0.5ex}}
\hspace{3.em} ({\l_{\bar{1}}} - 1)\,{\l_1}^{2/3}\,{\l_{\bar{1}}}^{1/3} = 
\z^2,
\label{TBA2Z3r}
\eea
with the eigenvalue equation. Since in the TBA equations (\ref{TBA1Z3r}) and 
(\ref{TBA2Z3r}) we still have two variables, $\l_1$ and $\l_{\bar{1}}$, 
the first step must be to reduce them to only one variable that represents 
their combined contribution to the free energy, like we did for unpolarized 
spinons. We have now that
\be
-\beta\,F = \sum_{j}\log\left[1+w_1(\zeta_j)^{-1}\right] +
2\sum_{j}\log\left[1+w_{\bar{1}}(\zeta_j)^{-1}\right] =
\sum_{j} \log\left[{\lambda_1}\,{\lambda_{\bar{1}}}^2\right]_j, \label{FZ3r}
\ee
since the particle of type ${\bar{1}}$ is a composite of 
two particles of type 1 and contributes double. 
Then we define a variable $\l = {\lambda_1}\,{\lambda_{\bar{1}}}^2$ to express 
the partition function in the factorized form 
$\prod_{j} \lambda ({\zeta}_j)$, corresponding to only one particle type. 
Now we can similarly reduce the two TBA equations to one equation for $\l$. 
It coincides with the eigenvalue equation (\ref{eigenZ3}). 

The central charge $c= 4/5$ can also be computed with the help of  
Eq.\ (\ref{eigenZ3}) but it involves the integral of its largest solution and 
has to be made numerically. However, the central charge can be 
evaluated exactly in this representation from the partition function 
(\ref{FZ3r}) and the TBA system (\ref{TBA1Z3r},\ref{TBA2Z3r}) in terms of the  
Rogers dilogarithm. Unlike the expression in the two particle representation 
(\ref{cZ3}) the calculation and the final expression are very complicated 
and are left for an appendix.

We can try to generalize the results above for the three-state Potts model 
to $\IZ_n$ parafermions. For example, the TBA equations in the one-particle 
representation of the $\IZ_4$ CFT turn out to be 
\bea
\left({{\lambda }_1 - 1} \right) \,
    {\sqrt{{{\lambda }_1}}}\,{{\lambda }_2}\,
    {\sqrt{{{\lambda }_3}}} = \zeta  ,\\
  \left({{\lambda }_2} - 1 \right) \,{{\lambda }_1}\,
    {{\lambda }_2}\,{{\lambda }_3} = {{\zeta }^2},\\
  \left( {{\lambda }_3} - 1 \right) \,
    {\sqrt{{{\lambda }_1}}}\,{{\lambda }_2}\,
    {\sqrt{{{\lambda }_3}}}
    = {{\zeta }^3},
\eea
and the free energy
\be
-\beta\,F = \sum_{j} \log\left[{\lambda_1}\,{\lambda_2}^2\,
{\lambda_3}^3\right]_j.
\label{FZ4r}
\ee
However, the algebraic calculation of the solution of these TBA equations is 
exceedingly difficult, even with algebraic computing programs. Nonetheless, 
one can solve them numerically. With a calculation of the solution in the 
interval $\z \in [0,1]$ at points separated by $0.1$ we have obtained the 
value $c=1.00008$ for the central charge, sufficiently accurate. 

\section{Discussion}

We have established the equivalence of the TBA approach and 
Schoutens' recursion matrix method in particular but representative 
examples. Since we have shown that both methods essentially consist 
of an approximate factorization of the partition function in 
the thermodynamic limit, we deem reasonable to conjecture that 
they are equivalent in general. However, in the examples in hand 
they seem to operate in slightly different ways. Both start with 
an exact expression of the partition function---say the CFT 
characters---in the form of an infinite sum but take different routes: 
The TBA approach uses the fractional statistics properties entailed 
by that expression, which directly lead to TBA equations and 
an approximate factorization of it. Schoutens' method 
demands first to find recursion matrix relations. In the examples 
studied here it has been done by inspection and, in our opinion, there is 
probably no systematic way to find recursion relations. 
Consequently, Schoutens' method appears more difficult to implement 
in practice. 

It is noteworthy that the problem of factorization of 
a CFT partition function can have several solutions with different 
physical interpretation. We have shown that for $\IZ_n$ parafermions 
there are two possible quasiparticle representations and two 
corresponding factorizations: The first is based on a set of 
$n-1$ fundamental quasi-particles whereas the second is based on 
only one, which generates the remaining $n-2$ as composites of two, 
three, etc., of it. The latter representation leads to a much more 
difficult TBA equation system and to new (and more difficult as well) 
dilogarithmic formulas for their central charges.

\appendix
\section{Calculation of the three-state Potts model central charge 
in the one-particle representation}

The central charge is $c = {\frac{6}{\pi^2}}(I_1 + 2\,I_2)$ with 
\bea
I_1 = \int _{0}^{1}\frac{d\z}{\z}\,\log (\l_1),\\
I_2 = \int _{0}^{1}\frac{d\z}{\z}\,\log (\l_{\bar{1}}).
\eea
To calculate $I_1$ we must first express $\z$ as function of only $\l_1$ 
eliminating $\l_2$ in the TBA system (\ref{TBA1Z3r},\ref{TBA2Z3r}). We obtain 
\be
{{\left( 1 - {{\lambda }_1} \right) }^3} + 
   {\z^3}\,{{\left(2 - {{\lambda }_1} \right) }^2}\,
    {{\lambda }_1} = 0,
\label{eql1}
\ee
and hence
\be
{d\z\over \z} = \left({1\over \l_1 -1}- {1\over 3\,\l_1} - 
{2 \over 3}\,{1\over \l_1 -2} \right)d\l_1.
\ee
Then
\be
I_1 = \int_0^\r {dx\over x}\log(1+x) - {1 \over 3}\,
\int_1^{1+\r} {d\l_1\over \l_1}\log(\l_1)- {2 \over 3}\,
\int_0^\r {dx\over x-1}\log(1+x),
\ee
where the change of integration variable $\l_1 = 1 + x$ has been used for 
the non-trivial first and third integrals. The integration limits are 
obtained by solving Eq.\ (\ref{eql1}) and are expressed in terms of 
$\r = \frac{-1 + {\sqrt{5}}}{2}$. 
 
The non-trivial integrals are of the type expressible as the Euler 
dilogarithm function:
$$
\int_0^\r {dx\over x}\log(1+x) = -Li_2(-\r),
$$
and, with the change of variable $x = 2\,u-1$,
$$
\int_0^\r {dx\over x-1}\log(1+x) = -\frac{\pi^2}{12} + 
\frac{{\log (2)}^2}{2} + 
     \log ({\frac{1 - \rho }{2}})\,\log (1 + \rho ) + 
     Li_2(2,{\frac{1 + \rho }{2}}).
$$  
So
\be
I_1 = -Li_2(-\r) 
- {1 \over 3}\,\frac{{{\log (1 + \rho)}^2}}{2} 
+ {{{\pi }^2} \over 18} - {{{\log (2)}^2}\over 3}  
     -{2\over 3}\,\log ({\frac{1 - \rho }{2}})\,\log (1 + \rho )  
     -{2\over 3}\,Li_2(2,{\frac{1 + \rho }{2}}).
\ee

We proceed analogously to calculate $I_2$. From the TBA system we get 
\be
{\z^6} + {{\left( 1 - {{\lambda }_{\bar{1}}} \right) }^3} - 
   2\,{\z^3}\,{{\left( -1 + {{\lambda }_{\bar{1}}} \right) }^2} = 0. 
\ee
Solving for $\z^3$
$$
\z^3 = \left( -1 + {{\lambda }_2} \right)
  \left( -1 \pm {\sqrt{-1 + {{\lambda }_2}}}\,
     {\sqrt{{{\lambda }_2}}} + {{\lambda }_2} \right)
$$
and selecting the positive solution we obtain 
\be
{d\z\over \z} = {d\l_2\over 3\,(\l_2 -1)} + {du\over 3\,u},
\ee
where we have introduced the new variable $u = -1 + {\sqrt{-1 + {{\lambda }_2}}}\,
     {\sqrt{{{\lambda }_2}}} + {{\lambda }_2}$ or
$$
\l_{\bar{1}} = {(1+u)^2\over 1+2\,u}.
$$
Then
\be
I_2 = {1 \over 3}\int_0^\r {dx\over x}\log(1+x) +
{2 \over 3}\int_0^{\r+1} {du\over u}\log(1+u) -
{1 \over 3}\int_0^{\r+1} {du\over u}\log(1+2\,u).
\ee
For the third integral we use the change of variable $v=2\,u$ and it becomes 
$$
\int_0^{2(\r+1)} {dv\over v}\log(1+v).
$$
Now we have two integrals equal to $Li_2(z)$ for $|z|>1$. 
We can transform them into standard dilogarithms with $|z|<1$ using 
the change of variable $u={y\over 1-y}$ and 
\be
\int_0^{z} {du\over u}\log(1+u) = 
-\int_0^{z\over z+1} {dy\over y\,(1-y)}\log(1-y).
\ee
They yield 
$$
\int_0^{\r+1} {du\over u}\log(1+u) = 2\,{{\log (\rho )}^2} + Li_2(\rho )
$$
and 
$$
\int_0^{2(\r+1)} {dv\over v}\log(1+v) = {\frac{9\,{{\log (\rho )}^2}}{2}} + 
  Li_2(2\,{{\rho }^2})
$$
after using the identity $\r+1 = 1/\r$ a number of times. Hence
\be
I_2 = {\frac{-{{\log (\rho )}^2}}{6}} - 
  {\frac{Li_2(-\rho )}{3}} + 
  {\frac{2\,Li_2(\rho )}{3}} - 
  {\frac{Li_2(2\,{{\rho }^2})}{3}}.
\ee

The contributions to the central charge from $\l_1$ and $\l_{\bar{1}}$, 
\bea
{\frac{6}{\pi^2}}\,I_1 = 0.421993347092241, \\
{\frac{12}{\pi^2}}\,I_2 = 0.378006652907758,
\eea
are trascendent numbers but their sum is 
\be
c = {\frac{6}{\pi^2}}(I_1 + 2\,I_2) = \frac{4}{5},
\ee
as expected. This can be proved as follows. In 
\bea
I_1 + 2\,I_2 = {\frac{{{\pi }^2}}{18}} - {\frac{{{\log (2)}^2}}{3}} - 
  {\frac{{{\log (\rho )}^2}}{3}} - 
  {\frac{2\,\log ({\frac{1 - \rho }{2}})\,\log (1 + \rho )}
    {3}} - {\frac{{{\log (1 + \rho )}^2}}{6}} \nonumber\\ \mbox{}- 
  {\frac{5\,Li_2(-\rho )}{3}} + 
  {\frac{4\,Li_2(\rho )}{3}} - 
  {\frac{2\,Li_2(2\,{{\rho }^2})}{3}} - 
  {\frac{2\,Li_2({\frac{1 + \rho }{2}})}{3}}
\eea
we can use the Euler identity
$$
Li_2(-x) = {\frac{Li_2({x^2})}{2}} -Li_2(x)
$$
to get
\bea
I_1 + 2\,I_2 = 
{\frac{{{\pi }^2}}{18}} - {\frac{{{\log (2)}^2}}{3}} - 
  {\frac{{{\log (\rho )}^2}}{3}} - 
  {\frac{2\,\log ({\frac{1 - \rho }{2}})\,\log (1 + \rho )}
    {3}} - {\frac{{{\log (1 + \rho )}^2}}{6}} \nonumber\\ \mbox{}+ 
  3\,Li_2(\rho ) - 
  {\frac{5\,Li_2({{\rho }^2})}{6}} - 
  {\frac{2\,Li_2(2\,{{\rho }^2})}{3}} - 
  {\frac{2\,Li_2({\frac{1 + \rho }{2}})}{3}}.
\eea
Now it is convenient to express it in terms of Rogers dilogarithms
$$
L(x) = Li_2(x) + \frac{1}{2}\,\log(x)\,\log(1-x),
$$
\bea
I_1 + 2\,I_2 =
{\frac{{{\pi }^2}}{18}} - {\frac{{{\log (2)}^2}}{3}} - 
  {\frac{3\,\log (1 - \rho )\,\log (\rho )}{2}} - 
  {\frac{{{\log (\rho )}^2}}{3}} + 
  {\frac{\log ({\frac{1 - \rho }{2}})\,
      \log ({\frac{1 + \rho }{2}})}{3}} \nonumber\\ \mbox{}- 
  {\frac{2\,\log ({\frac{1 - \rho }{2}})\,\log (1 + \rho )}
    {3}} - {\frac{{{\log (1 + \rho )}^2}}{6}} + 
  {\frac{\log (2\,{{\rho }^2})\,\log (1 - 2\,{{\rho }^2})}
    {3}} \nonumber\\
\mbox{}+ {\frac{5\,\log ({{\rho }^2})\,\log (1 - {{\rho }^2})}
    {12}} + 3\,L(\rho ) - 
  {\frac{5\,L({{\rho }^2})}{6}} - 
  {\frac{2\,L(2\,{{\rho }^2})}{3}} - 
  {\frac{2\,L({\frac{1 + \rho }{2}})}{3}}.
\label{I1I2}
\eea
The Rogers dilogarithms with arguments $\r$ and $\r^2$ can be easily 
calculated \cite{Lewin}. For the others, with the help of \cite{Lewin} 
$$
L(x) + L(1-x) = L(1) = {\frac{{{\pi }^2}}{6}},
$$
and the identities $1-2\,\r^2= \r^3$ and $1-\frac{1 + \rho }{2} = 
\frac{\r^2}{2}$, we obtain
\be
L(2\,{{\rho }^2}) + L({\frac{1 + \rho }{2}}) = 
2\,L(1) - \left[L(\r^3) + L(\frac{\r^2}{2})\right].
\ee
The last brackett can be reduced to computable Rogers dilogarithms 
\cite{Kiri}: 
\be
L(\r^3) + L(\frac{\r^2}{2}) = L(\r) + L(\r^2) - L(\frac{1}{2}).
\ee
It is another exercise in algebra to check that the contribution of 
the logarithmic terms in (\ref{I1I2}) vanishes. Finally,
\bea
I_1 + 2\,I_2 =  
{\frac{-{{\pi }^2}}{6}} - 
  {\frac{2\,L({\frac{1}{2}})}{3}} + 
  {\frac{11\,L(\rho )}{3}} - 
  {\frac{L({{\rho }^2})}{6}} \nonumber\\= 
{\frac{-{{\pi }^2}}{6}} - {\frac{2}{3}}\,{\frac{{{\pi }^2}}{12}} +
{\frac{11}{3}}\,{\frac{{{\pi }^2}}{10}} - {\frac{1}{6}}\,{\frac{{{\pi }^2}}{15}} 
= {\frac{4}{5}}\,{\frac{{{\pi }^2}}{6}}, 
\eea
using the known values \cite{Lewin}
$$
L({\frac{1}{2}}) = {\frac{{{\pi }^2}}{12}},\quad 
L(\rho ) = {\frac{{{\pi }^2}}{10}},\quad
L({{\rho }^2}) = {\frac{{{\pi }^2}}{15}}.
$$


\begin{thebibliography}{99}

\bibitem{DKKMM}S. Dasmahapatra, R. Kedem, T.R. Klassen, B.M. McCoy and E. Melzer, 
{\em Quasi-Particles, Conformal Field Theory, and $q$-Series}, 
Int. J. Mod. Phys. B7 (1993) 3617-3648, {\tt hep-th/9303013}
\bibitem{BerPaSer} F.D.M. Haldane, Z.N.C.\ Ha, J.C. Talstra, D. Bernard and 
V. Pasquier, Phys. Rev. Lett. 69 (1992) 2021;\\
D. Bernard, V. Pasquier and D. Serban, {\em Spinons in 
Conformal Field Theory}, Nucl. Phys. B428 (1994) 612-628, {\tt hep-th/9404050}
\bibitem{Haldane}F.D.M. Haldane, Phys. Rev. Lett. 67 (1991) 937
\bibitem{Wu}Y-S Wu, {\em Statistical Distribution for Generalized Ideal Gas of
Fractional-Statistics Particles}, Phys. Rev. Lett. 73 (1994) 922
\bibitem{Isakov}S.B. Isakov, S. Viefers, {\em Model of statistically coupled 
chiral fields on the circle}, Int. J. Mod. Phys. A12 (1997) 1895-1914, 
{\tt cond-mat/9701154}
\bibitem{YY}C.N. Yang and C.P. Yang, Jour. Math. Phys. 10 (1969) 1115
\bibitem{Schou}K. Schoutens, {\em Exclusion Statistics in 
Conformal Field Theory Spectra}, Phys. Rev. Lett. 79 (1997) 2608-2611, 
{\tt cond-mat/9706166}
\bibitem{ByFri}A.G. Bytsko and A. Fring, {\em Anyonic Interpretation of 
Virasoro Characters and the Thermodynamic Bethe Ansatz}, 
{\tt hep-th/9711113}
\bibitem{NaReTer}W. Nahm, A. Recknagel and M. Terhoeven, 
{\em Dilogarithm Identities in Conformal Field Theory},
Mod. Phys. Lett. A8 (1993) 1835-1848, 
{\tt hep-th/9211034}
\bibitem{Hal}F.D.M. Haldane, Phys. Rev. Lett. 60 (1988) 635
\bibitem{Shas}B.S. Shastry, Phys. Rev. Lett. 60 (1988) 639
\bibitem{BoLuSa}P. Bouwknegt, A. Ludwig and K. Schoutens, 
{\em Spinon Bases, Yangian Symmetry and Fermionic Representations of 
Virasoro Characters in Conformal Field Theory}, 
Phys. Lett. B338 (1994) 448-456, 
{\tt hep-th/9406020}
\bibitem{Lewin}L. Lewin, {\em Dilogarithm and associated functions}, North
Holland, Amsterdam, 1981
\bibitem{KKMM9211102}R. Kedem, T.R. Klassen, B.M. McCoy and E. Melzer, 
{\em Fermionic Quasi-Particle Representations for Characters of 
${(G^{(1)})_1 \times (G^{(1)})_1 \over (G^{(1)})_2}$}, 
Phys. Lett. B304 (1993) 263-270, {\tt hep-th/9211102}
\bibitem{DKMM93}S. Dasmahapatra, R. Kedem, B.M. McCoy and E. Melzer, 
{\em Virasoro Characters from Bethe Equations for the Critical Ferromagnetic 
Three-State Potts Model}, {\tt hep-th/9304150}
\bibitem{Zamo1}A.B. Zamolodchikov, Int. J. Mod. Phys. A 3 (1988) 743
\bibitem{Zamo2}Al.B. Zamolodchikov, Nucl. Phys. B 342 (1990) 695
\bibitem{Kiri}A.N. Kirillov, {\em Dilogarithm identities}, 
Prog.Theor.Phys.Suppl. 118 (1995) 61-142, {\tt hep-th/9408113}

\end{thebibliography}
\end{document}